\documentclass[11pt]{scrartcl}
\usepackage[pdftex]{graphicx}
\title{
Possible benefit of longer $L^*$ for detectors at ILC
}
\author{
Yasuhiro Sugimoto\footnote{Talk presented at the International Workshop
on Future Linear Colliders (LCWS2021),  15  -- 18 March, 2021. C21-03-15.1.}
\vspace{1cm}\\
High Energy Accelerator Research Organization (KEK) \\
Tsukuba, Ibaraki 305-0801, Japan
\date{}
}
\begin{document}
\maketitle
\begin{abstract}
In the present design of ILC, $L^*$, the distance between the interaction point (IP)
and the front surface of the final focusing magnet (QD0), is 4.1~m. The QD0
is supported from the detector platform by a cantilever type support
or from the detector itself.
This design could cause several difficulties in smooth operation of ILC
or in maintenance of detectors.
These difficulties could be mitigated by adopting longer $L^*$ optics.
In this article, we discuss possible benefit of longer $L^*$ for detectors. 

\end{abstract} 

%
\section{Introduction}
In the design of International Linear Collider (ILC)~\cite{adolphsen}, distance between 
the interaction point and the front surface of the final focusing quadrupole magnet, $L^*$,
is one of the essential parameters of the accelerator. 
In the baseline design, $L^*$ is 4.1~m~\cite{busser}.
The final focusing quadrupole magnets (QD0) locate inside of the detector,
and supported from the detector itself (SiD) or from a cantilever 
support structure (ILD)~\cite{behnke}.
In both cases, QD0 is supported from the movable platform
on which the detector is placed.  In this configuration, 
there are risks of;
\begin{itemize}
  \item QD0 vibration caused by the  detector,
  \item poor position repeatability of QD0 and longer time for machine tuning
  after push-pull operation,
  \item difficulty in opening detector endcap.
\end{itemize}

After publication of the ILC TDR~\cite{adolphsen}, longer $L^*$ optics has been 
proposed~\cite{okugi, plassardm, plassardd}
and seriously studied for CLIC. In the recent CLIC design $L^*=6$~m is adopted.
If we could adopt the similar optics for ILC, the risks listed above would be
greatly mitigated.

\section{ILD endcap opening}
\begin{figure}
\centering
\includegraphics[scale=0.7]{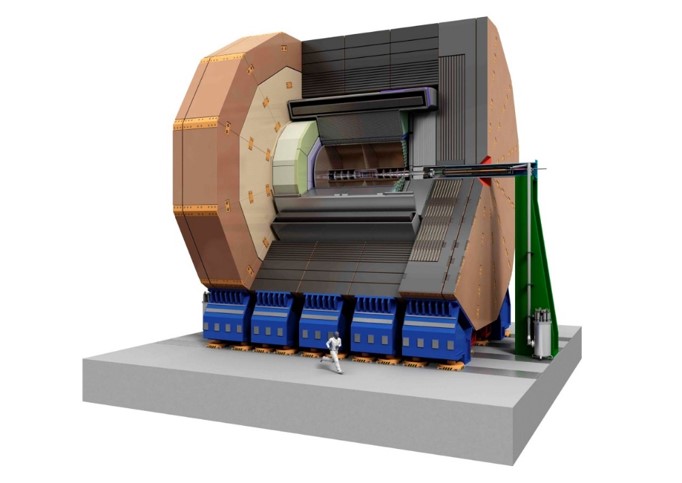}
\includegraphics[scale=0.6]{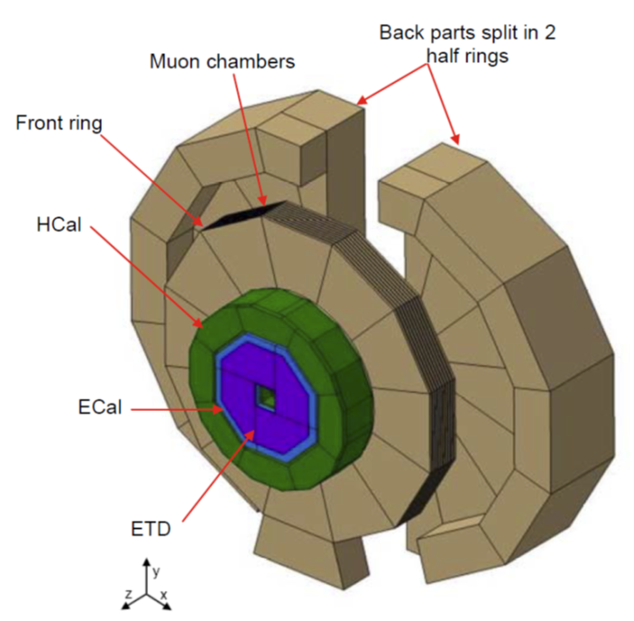}
\caption{Cut view of ILD detector (left) and structure of its endcap (right). 
The gray part below the detector in the left figure is the movable platform.}
\label{ILD}
\end{figure}

The cut view of ILD detector is schematically shown in Figure~\ref{ILD} (left).
QD0 is supported from a cantilever structure standing on the
movable platform. 
Because the pillar of this structure standing on the movable platform 
locates just behind the detector endcap,
the rear part of the endcap has to be split as shown in Figure~\ref{ILD} (right)
in order to clear the pillar~\cite{sinram}. 
About 1.2~m gap can be obtained between the endcap and the barrel part of the
detector by splitting the rear part of the endcap. 
However, if more gap is needed, for example, to uninstall the central tracker,
the QD0 support structure has to be withdrawn by few meters. 
The split endcap requires additional engineering challenges.
Because each part of the split endcap needs own support legs, 
more space on the floor is necessary to make it stable against earth quake.
Lowering method from surface assembly hall to the underground detector hall
using a gantry crane could be more complicated.

\section{ILD with $L^* = 6$~m}
\begin{figure}
\centering
\includegraphics[scale=0.47]{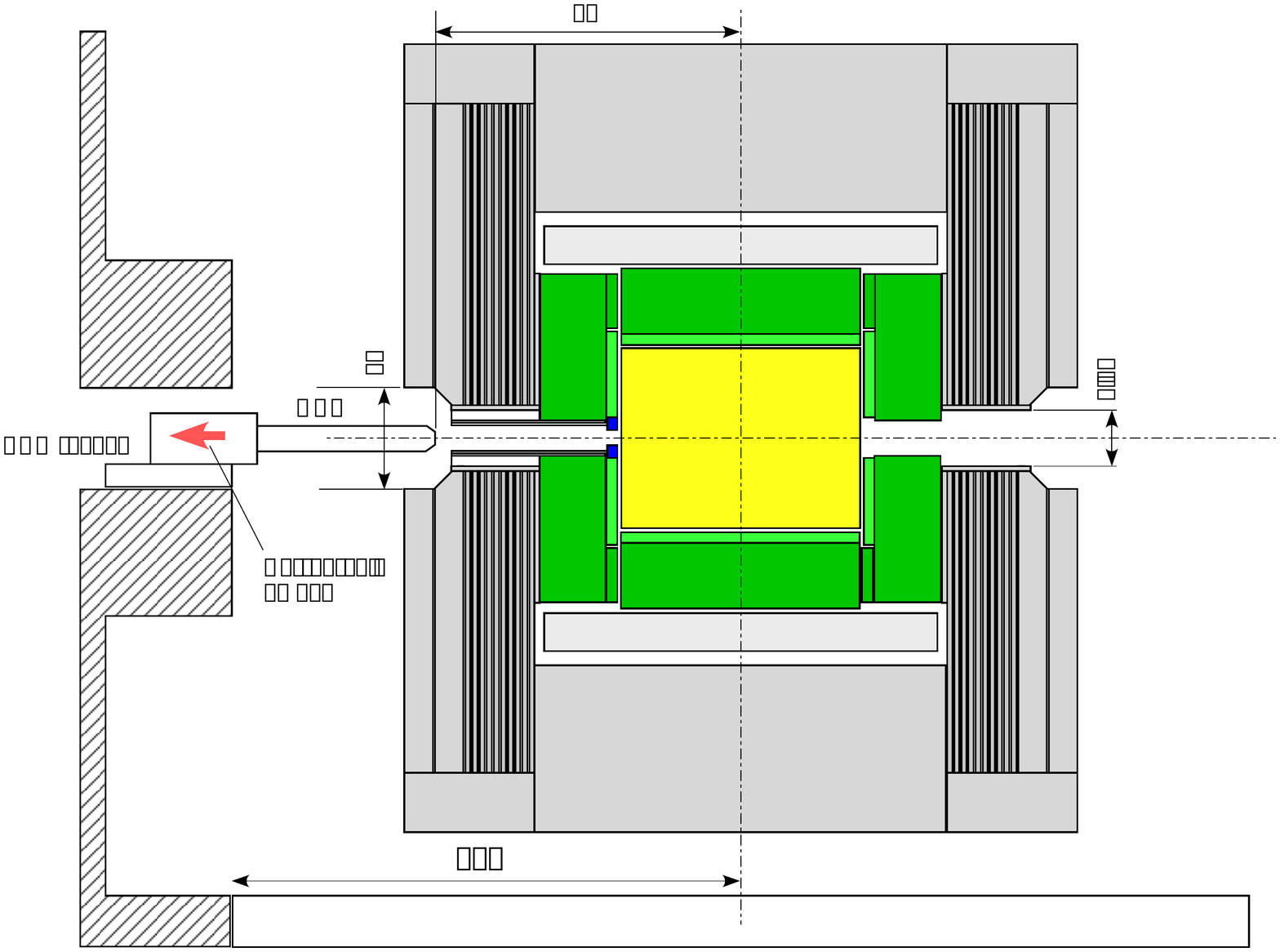}
\includegraphics[scale=0.47]{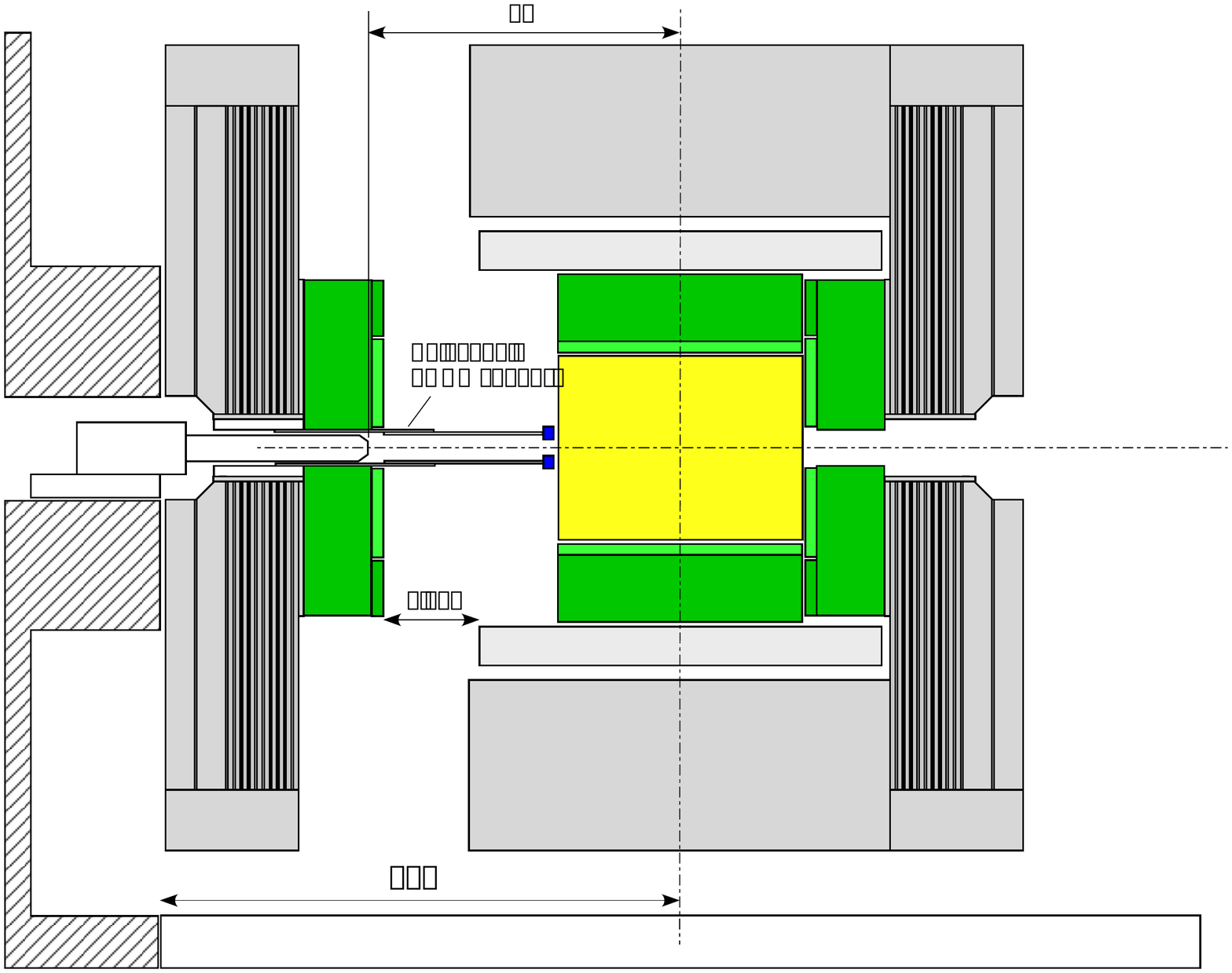}
\caption{ILD with $L^*=6$~m when the endcap is closed (top) and open(bottom).}
\label{ILD6m}
\end{figure}

If $L^*=6$~m option can be adopted for ILC, it would bring a large benefit for
detectors. In this case, QD0 can be supported from the floor of the beam tunnel
as shown in Figure~\ref{ILD6m}, and we can get rid of QD0 support pillar 
standing on the movable platform. In order to make the push-pull operation possible,
QD0 should be retractable using a sliding support 
like QCS of SuperKEKB~\cite{superKEKB}.
Because the QD0 support pillar does not exist in this design, 
the endcap of ILD does not have to be split.  
The endcap can be easily opened  at the 
maintenance position when a barrel part detector has to be uninstalled.

Another merit of longer $L^*$ option could be reduction of beam background
due to back-scattering of $e^+/e^-$ pair-background from the small angle
calorimeter (BCAL) or QD0. The small angle calorimeter, BCAL,
is supposed to be placed near the front surface of QD0. If $L^*$ becomes longer,
BCAL can be placed farther from the interaction point where the magnetic field
caused by the detector solenoid is much weaker. Because the back-scattered 
$e^+/e^-$ goes back to the central detector region spiraling along the magnetic 
field line,
the number of back-scattered particles reaching the central detectors would 
largely reduced if the source is located at the weak magnetic field region.
In the ILD baseline design, an anti-DID (Detector-Integrated-Dipole) is 
used to guide the dense part of pair background particles 
from the interaction point to the beam holes
of the BCAL and QD0 along the magnetic field line. 
On the other hand, if the magnetic field
around the BCAL and QD0 is weak, this anti-DID may not work, and increase
number of background particles hitting the BCAL and QD0.
We need a detailed simulation study    of the background to see if
the back-scattered background can be reduced. If the result is that
the anti-DID does not work, but the backscattered background is reduced,
then we can discard the anti-DID, and reduce the cost,  construction period,
and risks.

In addition to the position of BCAL, some modification of ILD is needed.
For the push-pull operation, the beam pipe has to be disconnected
at $z\sim 6$~m. In order to make access to flanges at $z\sim 6$~m possible,
the hole of the endcap has to be enlarged there. The remote vacuum connection
(RVC) adopted for SuperKEKB could be another solution.

The longer $L^*$ option is also preferable from the viewpoint of
smooth operation of ILC.  QD0 supported from the floor of the beam tunnel
is much more stable against the vibration than that supported from the support pillar 
standing on the movable detector platform. 
Much better position repeatability of QD0 can be achieved after push-pull operation.
In the case of the QD0 support from the movable platform,  
the anticipated repeatability is $\sim 0.5$~mm.
On the other hand, in the case of the sliding support on the floor of the beam tunnel,  
we can expect to achieve the position 
repeatability of few tens of $\mu$m from the experience in SuperKEKB. 
The better repeatability would make the time for machine tuning after push-pull operation
much faster, which could compensate possible luminosity reduction in longer $L^*$ optics.

 \section{Summary} 
 The design of QD0 support in TDR has possible risks in stable and smooth operation
 of ILC, as well as difficulties in detector design and maintenance. 
 The QD0 support from the beam tunnel with longer $L^*$ could solve
 these issues. 
 The longer $L^*$ option might also contribute to
 reduction of cost and possibly construction period of ILD detector.
 Therefore, this option should be seriously studied during early phase
 of Pre-Lab period, both from the accelerator and detector point of view.

\begin{footnotesize}

\end{footnotesize}


\end{document}